\newcommand{\etac}{\eta_{c}}

\newcommand{\jpsi}{J/\psi}

\newcommand{\pip}{\pi^+}
\newcommand{\pin}{\pi^-}
\newcommand{\pio}{\pi^0}
\newcommand{\ppb}{p\bar{p}}

\newcommand{\jpsito}{J/\psi \rightarrow }

\newcommand{\beq}{\begin{equation}}
\newcommand{\eeq}{\end{equation}}
\newcommand{\beqn}{\begin{eqnarray}}
\newcommand{\eeqn}{\end{eqnarray}}
\newcommand{\beqns}{\begin{eqnarray*}}
\newcommand{\eeqns}{\end{eqnarray*}}
\newcommand{\bfg}{\begin{figure}}
\newcommand{\efg}{\end{figure}}
\newcommand{\bitm}{\begin{itemize}}
\newcommand{\eitm}{\end{itemize}}
\newcommand{\bnum}{\begin{enumerate}}
\newcommand{\enum}{\end{enumerate}}
\newcommand{\btbl}{\begin{table}}
\newcommand{\etbl}{\end{table}}
\newcommand{\btbu}{\begin{tabular}}
\newcommand{\etbu}{\end{tabular}}

\newcommand{\etap}{\eta^{\prime}}
\newcommand{\rhoo}{\rho^0}

\newcommand{\be}{\begin{enumerate}}
\newcommand{\ee}{\end{enumerate}}
\newcommand{\bi}{\begin{itemize}}
\newcommand{\ei}{\end{itemize}}
\newcommand{\rar}{\rightarrow}
\newcommand{\ppp}{\pi^+\pi^-\pi^0}

\documentclass[preprint,showpacs,amsmath,amssymb]{revtex4}
\usepackage{rotating,epsfig,graphicx}
\usepackage{flafter}
\include{def-com}
\normalsize
\begin{document}
\title{\boldmath \bf Measurement of the Branching Fractions for $J/\psi$
$\rar$ $p\bar{p}\eta$ and $p\bar{p}\etap$ }
\author{
M.~Ablikim$^{1}$,              J.~Z.~Bai$^{1}$,   Y.~Bai$^{1}$,
Y.~Ban$^{11}$,
X.~Cai$^{1}$,                  H.~F.~Chen$^{16}$,
H.~S.~Chen$^{1}$,              H.~X.~Chen$^{1}$, J.~C.~Chen$^{1}$,
Jin~Chen$^{1}$
% ...
\\
\vspace{0.2cm}
(BES Collaboration)\\
\vspace{0.2cm}
{\it
$^{1}$ Institute of High Energy Physics, Beijing 100049, People's Republic of China\\
% ...
$^{11}$ Peking University, Beijing 100871, People's Republic of China\\
% ...
$^{16}$ University of Science and Technology of China, Hefei 230026, 
People's Republic of China\\
% ...
}}

\begin{abstract}
  Using 58$\times 10^{6}$ $\jpsi$ events collected with the Beijing
  Spectrometer (BESII) at the Beijing Electron Positron Collider
  (BEPC), the branching fractions of $\jpsi$ to $p\bar{p}\eta$ and
  $p\bar{p}\etap$ are determined. The ratio
  $\frac{\Gamma(\jpsi\rar\ppb\eta)}{\Gamma(\jpsi\rar\ppb)}$ obtained
  by this analysis agrees with expectations based on soft-pion theorem
  calculations.
\end{abstract}
\maketitle

\section{\boldmath Introduction}   \label{introd}
The $\jpsi$ meson has hadronic, electromagnetic,
and radiative decays to light hadrons, and a radiative transition to
the $\etac$. In Ref.~\cite{PCX}, direct hadronic, electromagnetic and
radiative decays are estimated to account for 69.2$\%$, 13.4$\%$, and
4.3$\%$, respectively, of all $\jpsi$ decays. However, individual
exclusive $\jpsi$ decays are more difficult to analyze quantitatively
in QCD. To date, two-body decay modes such as $\jpsito
B_{8}\bar{B_{8}}$ or $P_{9}V_{9}$, where $B_{8}$, $P_{9}$ and $V_{9}$
refer to baryon octet, pseudoscalar nonet, and vector nonet particle,
respectively, have been studied with some success using an effective
model, and other similar methods~\cite{SPIT}. 

Studies of three-body decays of $\jpsi$ are a natural extension of
studies of two-body decays. Since most $\jpsi$ decays proceed via
two-body intermediate states, including wide resonances, it is hard to
experimentally extract the non-resonant three-body
contribution~\cite{JPDC}.  Specific models based on proton and $N^{*}$
pole diagrams have been introduced to deal with these
problems~\cite{SPIT}.  In the calculation, the soft-pion
theorem~\cite{LABW} has been applied to the decay $\jpsito \ppb\pio$
successfully. This method has also been used for $\jpsito \ppb\eta$ and
$\jpsito\ppb\etap$ decays~\cite{SPIT}.

 This paper reports measurement of the
branching fractions for $p\bar{p}\eta$ and $p\bar{p}\etap$, and tests
of the soft-pion theorem for $\jpsito \ppb\eta$, which
states~\cite{SPIT}:
\begin{eqnarray*}
\frac{\Gamma(\jpsi\rar\ppb\eta)}{\Gamma(\jpsi\rar\ppb)}\simeq 0.64\pm0.52 .
\end{eqnarray*}

\section{\boldmath The BES detector and Monte Carlo simulation}  \label{BESD}
BESII is a conventional solenoidal magnet detector that is described
in detail in Refs.~\cite{JZB}. A 12-layer vertex chamber (VC)
surrounding the beam pipe provides trigger and track information. A
forty-layer main drift chamber (MDC), located radially outside the VC,
provides trajectory and energy loss ($dE/dx$) information for tracks
over $85\%$ of the total solid angle.  The momentum resolution is
$\sigma _p/p = 0.017 \sqrt{1+p^2}$ ($p$ in $\hbox{\rm GeV}/c$), and
the $dE/dx$ resolution for hadron tracks is $\sim 8\%$.  An array of
48 scintillation counters surrounding the MDC measures the
time-of-flight (TOF) of tracks with a resolution of $\sim 200$ ps for
hadrons.  Radially outside the TOF system is a 12 radiation length,
lead-gas barrel shower counter (BSC).  This measures the energies of
electrons and photons over $\sim 80\%$ of the total solid angle with
an energy resolution of $\sigma_E/E=22\%/\sqrt{E}$ ($E$ in GeV).
Outside of the solenoidal coil, which provides a 0.4~Tesla magnetic
field over the tracking volume, is an iron flux return that is
instrumented with three double layers of counters that identify muons
of momentum greater than 0.5~GeV/$c$.

In the analysis, a GEANT3-based Monte Carlo (MC) simulation program
(SIMBES)~\cite{GEANT} with detailed consideration of detector
performance is used. The consistency between data and MC has been
validated using many high purity physics channels~\cite{NIM}.

In this analysis, the detection efficiency for each decay mode is
determined by a MC simulation that takes into account the angular
distributions. For $\jpsi\rar\ppb\eta$, the angle ($\theta$) between
the directions of $e^{+}$ and $p$ in the laboratory frame is generated
according to $1+\alpha\cdot\cos^{2}\theta$ distribution, where
$\alpha$ is obtained by fitting the data from $\jpsi\rar\ppb\eta$. A
uniform phase space distribution is used for $\jpsi$ decaying into
$\ppb\etap$.

\section{\boldmath General Selection Criteria}
Candidate events are required to satisfy the following common
selection criteria:
\subsection{Charged track selection}
Each charged track must: (1) have a good helix fit in order to ensure
a correct error matrix in the kinematic fit; (2) originate from the
interaction region, $\sqrt{V^{2}_{x}+V^{2}_{y}}<2$ cm and $|V_{z}|<20$
cm, where $V_{x}$, $V_{y}$, and $V_{z}$ are the $x$, $y$ and $z$
coordinates of the point of closest approach of the track to the beam
axis; (3) have a transverse momentum greater than 70 MeV/$c$; and (4)
have $|\cos \theta|\le 0.80$, where $\theta$ is the polar angle of the
track.
\subsection{Photon selection} 
A neutral cluster in the BSC is assumed to be a photon candidate if
the following requirements are satisfied: (1) the energy deposited in
the BSC is greater than 0.05 GeV; (2) energy is deposited in more than
one layer; (3) the angle between the direction of photon emission and
the direction of shower development is less than $30^{\circ}$; and (4)
the angle between the photon and the nearest charged track is greater
than $5^{\circ}$ (if the charged track is an antiproton, the angle is
required to be great than $25^{\circ}$).
\subsection{Particle Identification (PID)}
For each charged track in an event, $\chi^{2}_{PID}(i)$ is determined
using both $dE/dx$ and TOF information:
\begin{center}
$\chi^{2}_{PID}(i)=\chi^{2}_{dE/dx}(i)+\chi^{2}_{TOF}(i)$, \\
\end{center}
where $i$ corresponds to the particle hypothesis. A charged track is
identified as a pion if $\chi^{2}_{PID}$ for the $\pi$ hypothesis is
less than those for the kaon and proton hypotheses. For $p$ or
$\bar{p}$ identification, the same method is used. In this analysis,
all charged tracks are required to be positively identified.

\section{\boldmath Analysis of $\jpsi \rar \ppb\eta$}
The decay modes for the $\jpsi \rar\ppb\eta$ measurement are $\eta
\rar \gamma \gamma$ and $\eta \rar \ppp$. The use of different decay
modes allows us to cross check our measurements, as well as to obtain
higher statistical precision.
\begin{figure}[htpb]
\vspace{-2.0cm}
\begin{center}
\psfig{file=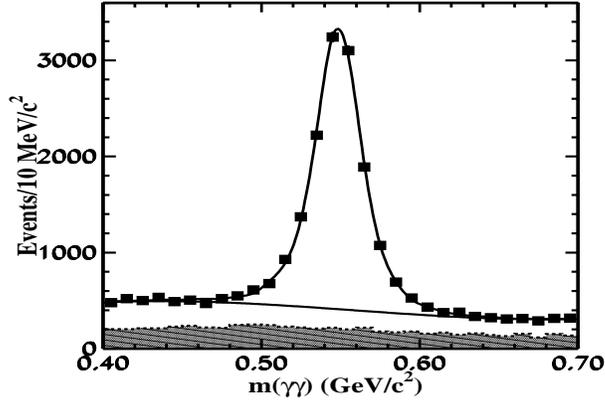,width=8cm,height=8cm,angle=0}
\vspace{-0.5cm}
\caption{The two-photon invariant mass distribution for $\jpsi \rar
  \ppb\gamma\gamma$ candidate events. Data are represented by
  rectangles; the error bars are too small to be seen. The curves are
  the results of the fit described in the text. The shaded part is
  background from MC simulation.}
\label{mggfit}
\end{center}
\end{figure}
\subsection{\boldmath $\eta \rar \gamma\gamma$}
Events with two charged tracks and two photons are selected. A
four-constraint (4C) kinematic fit is performed to the hypothesis
$\jpsi \rar \ppb \gamma\gamma$. For events with more than two photons,
all combinations are tried, and the combination with the smallest
$\chi^{2}$ is retained.  $\chi^{2}_{\gamma\gamma\ppb}$ is required to
be less than 20.

The $\gamma\gamma$ invariant mass $(m_{\gamma\gamma})$ distribution
for selected events is shown in Fig.~\ref{mggfit}. A peak around the
$\eta$ mass is evident. The curves in the figure indicate the best fit
to the signal and background. The shaded part is the background
estimated from a MC simulation of inclusive $\jpsi$
events~\cite{LUND}. The main background comes from $\jpsi \rar
\ppb\pio\pio$ and $\Sigma^{+}\bar{\Sigma}^{-}$.
By fitting the $\eta$ signal with a MC-simulated signal
histogram plus a third order polynomial background function, the
number of $\eta$ signal events is determined to be $(12220\pm149)$.

For the signal MC simulation, the events are generated with a proton
angle distribution of $1+\alpha\cos^{2}\theta$, where $\alpha$ is
taken to be -0.6185 in order to describe the data. In the decay,
intermediate resonances, N(1440), N(1535), N(1650), and N(1800) and
antiparticles, with fractional contribution of $(8\pm4)$\%,
$(56\pm15)$\%, $(24^{+5}_{-15})$\%, and $(12\pm7)$\%~\cite{LHBD},
respectively, are included. The resulting detection efficiency for
$\jpsi\rar\ppb\eta$ $(\eta\rar\gamma\gamma)$ is determined to be
$28.70$\%. The $\ppb\eta$ branching fraction, calculated
using
\begin{eqnarray*}
B(\jpsi\rar\ppb\eta)=\frac{N_{obs}}{\epsilon\cdot N_{\jpsi}\cdot
B(\eta\rar 2\gamma)\cdot f_{1}},
\end{eqnarray*}
 is $(1.93\pm0.02)\times 10^{-3}$, where the error is statistical
only.  Here $N_{obs}$ represents the number of observed events,
$\epsilon$ is the detection efficiency for $\jpsi\rar\ppb\eta (\eta\to
2\gamma)$, $f_{1} = 0.9739$ is the efficiency correction factor (see
Section~\ref{SYSE}), and $N_{\jpsi}$ is the total number of $\jpsi$
events.
\begin{figure}[htpb]
\vspace{-2.0cm}
\begin{center}
\psfig{file=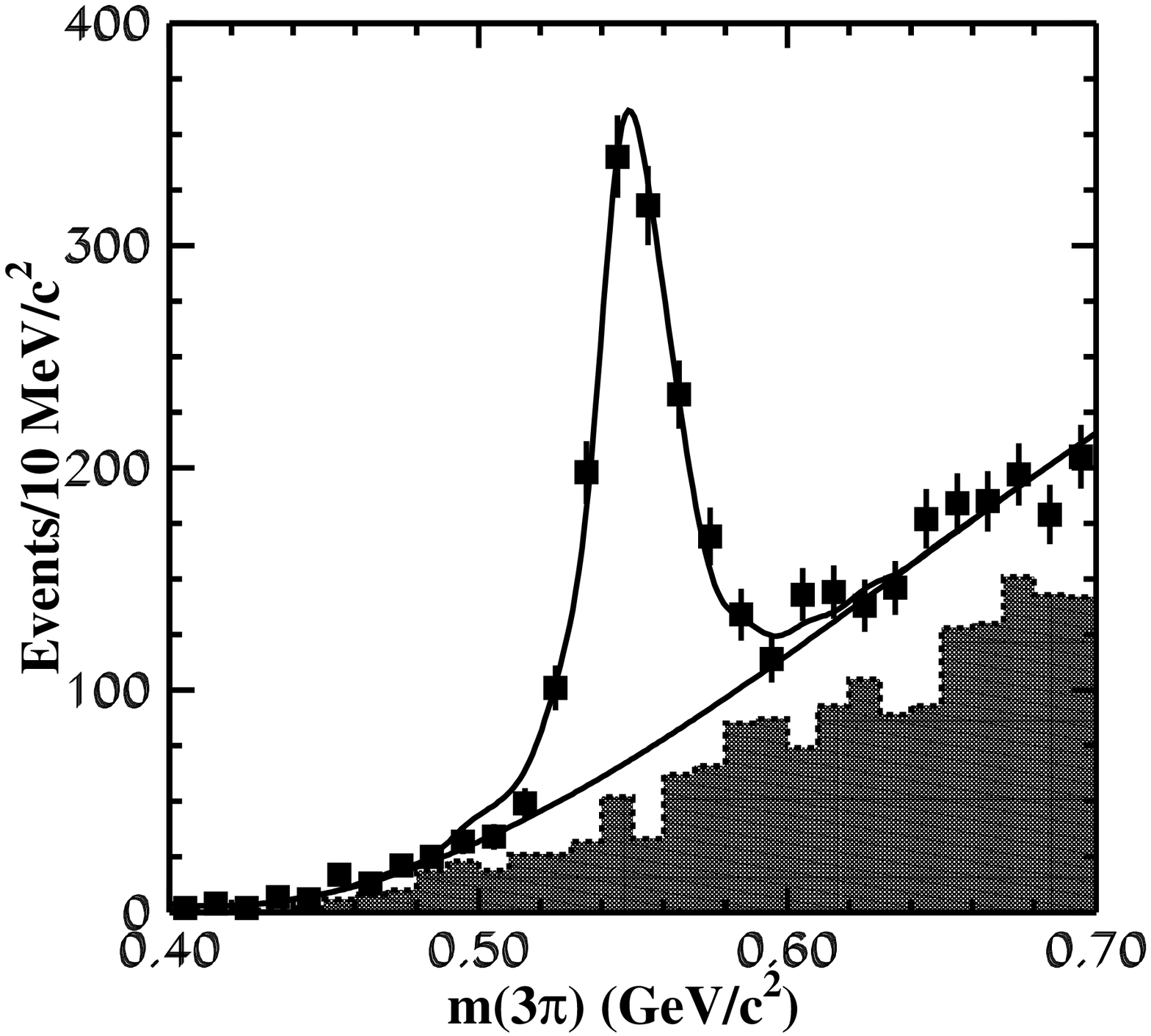,width=8cm,height=8cm,angle=0}
\vspace{-0.5cm}
\caption{The $\ppp$ invariant mass distribution for $\jpsi \rar
\ppb\ppp$ candidate events. The curves are
results of the fit described in the text. The shaded part is
background from MC simulation.} \label{m3pifit}
\end{center}
\end{figure}

\subsection{\boldmath $\eta\rar\pip\pin\pio$}
Similar to the above analysis, events with four charged tracks and two
photons are selected. A 4C kinematic fit is performed to the
$\jpsi\rar\ppb\pip\pin\gamma\gamma$ hypothesis, and the
$\chi^{2}_{\gamma\gamma\ppb\pip\pin}$ value is required to be less
than 20. In order to suppress multi-photon backgrounds, the number of
photons is required to be two. The invariant mass of the
$\gamma\gamma$ is required to be between 0.095 and 0.175 GeV/$c^{2}$.

The $\ppp$ invariant mass $(m_{\ppp})$ distribution is shown in
Fig.~\ref{m3pifit}, where a peak at the $\eta$ mass is observed. The
curves in the figure are the results of a fit to the signal and
background. The shaded part is background estimated
from MC simulation of inclusive $\jpsi$ decay events~\cite{LUND}. Here
the main background comes from $\jpsi \rar \ppb\ppp$ and
$\ppb\pip\pin\gamma$ decays.
By fitting the distribution with a MC simulated signal histogram plus
a third order polynomial background function, $(954\pm45)$ signal
events are obtained. Similar to the $\eta \rar 2 \gamma$ decay,
contributions from the baryon excited states N(1440), N(1535),
N(1650), and N(1800), as well as their anti-particles~\cite{LHBD}, are
considered. The detection efficiency of $\jpsi\rar\ppb\eta$
$(\eta\rar\ppp$, $\pio\rar\gamma\gamma)$ is determined to be
$4.20$\%. The branching fraction is determined from the
calculation
\begin{eqnarray*}
B(\jpsi\rar\ppb\eta)&=&\frac{N_{obs}}{\epsilon\cdot N_{\jpsi}\cdot
B(\eta\rar \ppp)}\\
&\cdot& \frac{1}{B(\pio\rar \gamma\gamma)\cdot f_{2}},
\end{eqnarray*}
where $f_{2} = 0.9582$ is a correction factor for the efficiency that is
described below in Section~\ref{SYSE}.  We
determine a branching fraction for $\jpsi\rar\ppb\eta$ of
$(1.83\pm0.09)\times 10^{-3}$, where the error is statistical only.

\begin{figure}[htpb]
\vspace{-2.0cm}
\begin{center}
\psfig{file=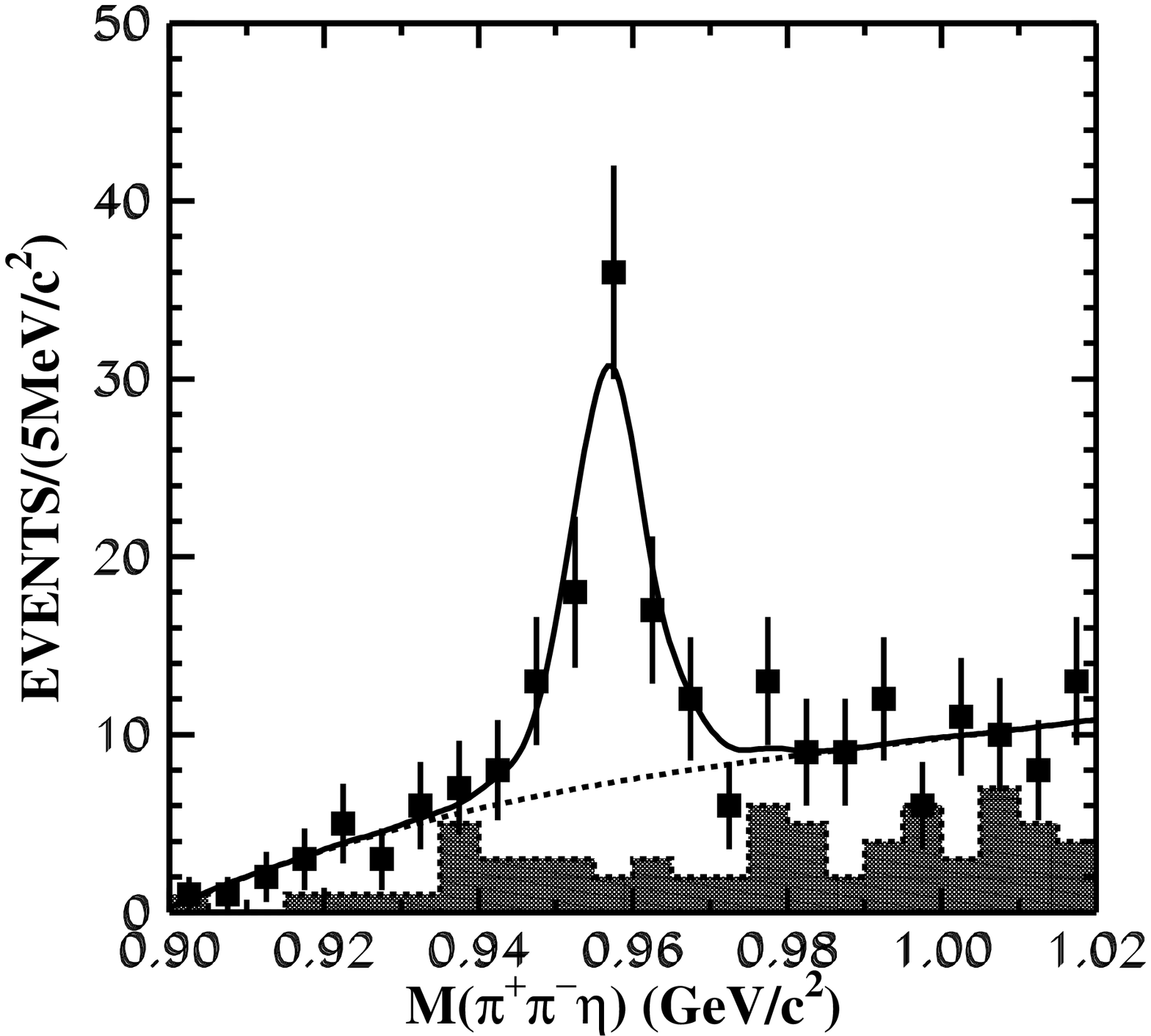,width=8cm,height=8cm,angle=0}
\vspace{-0.5cm}
\caption{The $\pip\pin\eta$ invariant mass distribution for $\jpsi
  \rar \ppb\pip\pin\eta$ candidate events. The curves are results of
  the fit described in the text. The shaded part is background from MC
  simulation.}
\label{metp1fit}
\end{center}
\end{figure}

\section{\boldmath Analysis of $\jpsi\rar\ppb\etap$}
There are three main decay modes of the $\etap$:
$\etap \rar \pip\pin\eta$, $\etap\rar\gamma\rhoo$ and
$\etap\rar\pio\pio\eta$. Here the first two decay modes are used.

\subsection{\boldmath $\etap\rar\pip\pin\eta$, $\eta\rar\gamma\gamma$}
In the search for $\etap\rar\pip\pin\eta$ decays, events with four
charged tracks and two photons are selected. A five-constraint (5C)
kinematic fit is performed to the hypothesis of
$\jpsi\rar\ppb\pip\pin\gamma\gamma$, in which the $2\gamma$ invariant
mass is constrained to equal the $\eta$ mass, and the
$\chi^{2}_{\gamma\gamma\ppb\pip\pin}$ value is required to be less
than 20.

The $\pip\pin\eta$ invariant mass $(m_{\pip\pin\eta})$ distribution
for events that survive the selection criteria is shown in
Fig.~\ref{metp1fit}. A clear $\etap$ signal is observed. The curves in
the figure are the best fit to the signal and background. The shaded
part is background estimated from MC simulation of
inclusive $\jpsi$ decay events~\cite{LUND}.  The main background comes
from $\jpsi \rar \Delta^{+}\bar{\Delta}^{-}\eta$, and
$\Delta^{0}\bar{\Delta}^{0}\eta$ decays.
By fitting the distribution with a MC simulated signal histogram plus
a third order polynomial background function, a signal yield of
$(65\pm12)$ events is observed. According to a MC simulation, in which
the events are generated with uniform phase space, the detection
efficiency of $\jpsi\rar\ppb\etap$ $(\etap\rar\pip\pin\eta$,
$\eta\rar\gamma\gamma)$ is $3.38$\%. The effect of
intermediate resonances is considered as a source of systematic error.
Using
\begin{eqnarray*}
B(\jpsi\rar\ppb\etap)&=&\frac{N_{obs}}{\epsilon\cdot N_{\jpsi}\cdot
B(\etap\rar \pip\pin\eta)}\\
&\cdot& \frac{1}{B(\eta\rar \gamma\gamma)\cdot f_{3}}
\end{eqnarray*}
with $f_{3} = 0.8228$ being the efficiency correction factor (see
Section~\ref{SYSE}), we determine the branching fraction for
$\jpsi\rar\ppb\etap$ to be $(2.31\pm0.43)\times 10^{-4}$, where the
error is statistical only.

\begin{figure}[htpb]
\vspace{-2.0cm}
\begin{center}
\psfig{file=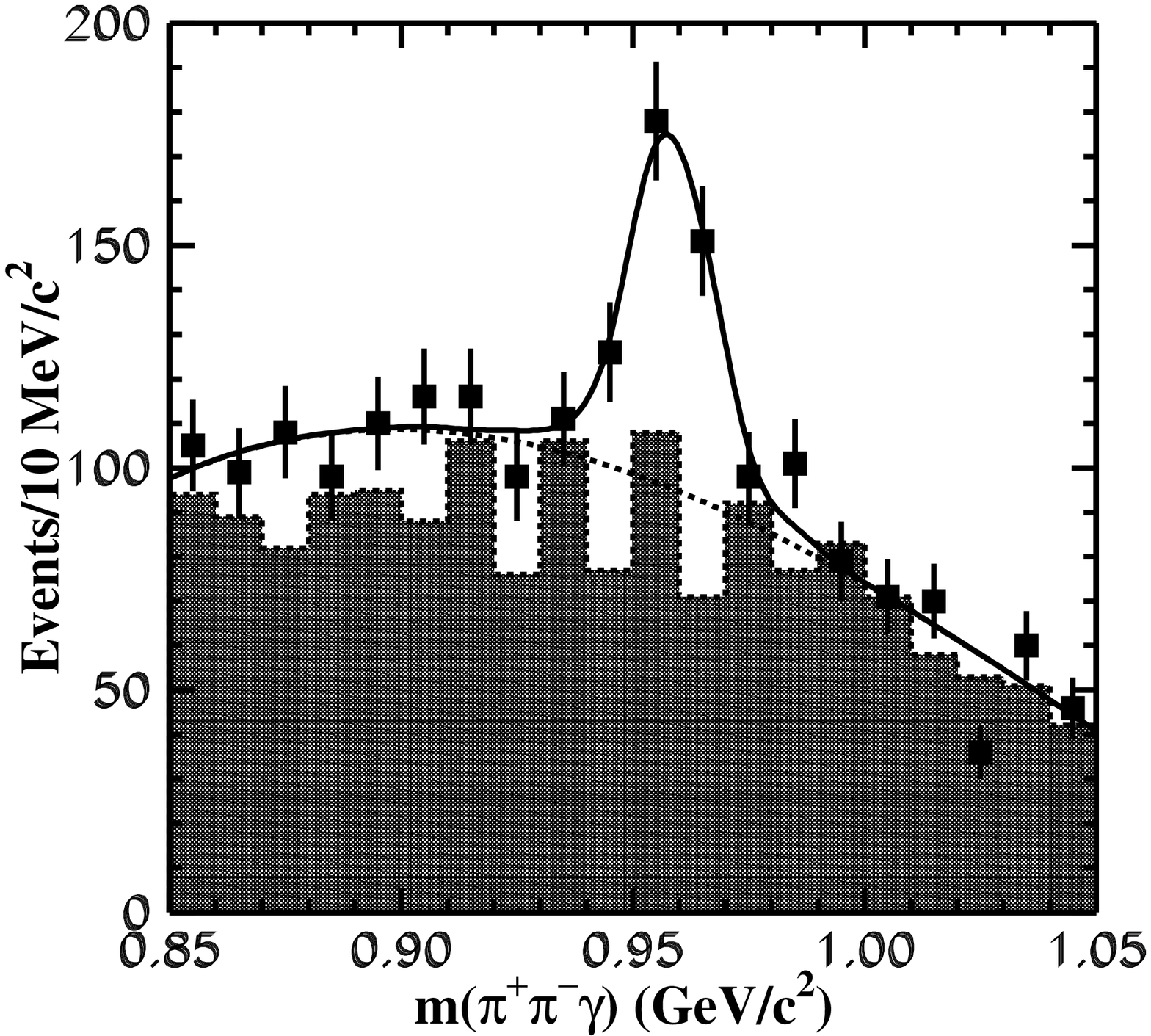,width=8cm,height=8cm,angle=0}
\vspace{-0.5cm}
\caption{The $\gamma \pip\pin$ invariant mass distribution for
  $\jpsi\rar\ppb\gamma\pip\pin$ candidate events. The curves are
  results of the fit described in the text. The shaded part is
  background from MC simulation.} \label{metp2fit}
\end{center}
\end{figure}

\subsection{\boldmath $\etap\rar\gamma\rhoo,\rhoo\rar\pip\pin$}
In order to select $\etap\rar\gamma\rhoo$, a 4C kinematic fit is
performed under the hypothesis of $\jpsi\rar\ppb\pip\pin\gamma$. The
$\chi^{2}_{\gamma\ppb\pip\pin}$ value is required to be less than 20.
To ensure the events are from $\gamma\rho^{0}$, a
$|m_{\pip\pin}-m_{\rho}|<0.20$ GeV/$c^{2}$ requirement is imposed,
where $m_{\pip\pin}$ is the $\pip\pin$ invariant mass, and $m_{\rho}$
is the $\rho$ mass. In order to exclude the background from $\jpsi\rar
p\bar{p}\pip\pin$, it is required that the invariant mass of the four
charged tracks is less than 3.02 GeV/$c^{2}$.

The $\gamma\rhoo$ invariant mass $(m_{\gamma\rhoo})$ distribution for
selected events, where a clear $\etap$ signal is observed, is shown
in Fig.~\ref{metp2fit}. The curves in the figure are the best fit to
the signal and background. The shaded part is the background estimated
from MC simulation of inclusive $\jpsi$ decay events~\cite{LUND}. 
The main background comes
from $\jpsi \rar \ppb\pip\pin\gamma$,
${\Delta}^{++}\bar{\Delta}^{--}\pio$ and $\ppb\ppp$ decays.  By
fitting the $m_{\gamma\pip\pin}$ distribution with a MC simulated
signal shape and a third order polynomial background function, we
determine the number of $\etap$ signal events to be $(200\pm29)$.  The
detection efficiency for $\jpsi\rar\ppb\etap$ $(\etap\rar \gamma\rhoo)$
is determined to be $7.48$\%, assuming phase space
production, where the $\pip\pin$ mass distribution is generated
according to measurements
from$\jpsi\rar\phi\etap,\etap\rar\gamma\pip\pin$~\cite{FPRD71}.  Using
\begin{eqnarray*}
B(\jpsi\rar\ppb\etap)&=&\frac{N_{obs}}{\epsilon\cdot
N_{\jpsi}\cdot B(\etap\rar \gamma\rhoo)\cdot f_{4}}
\end{eqnarray*}
with the $f_{4}$ correction factor of 0.8522 (see Section~\ref{SYSE}).
The resulting branching fraction for $\jpsi\rar\ppb\etap$
is $(1.85\pm0.27)\times 10^{-4}$, where the error is statistical only.

\begin{table*}
\begin{center}
\caption{ Numbers used in the calculations of branching fractions and results.}
\vspace{0.5cm}
\label{TABFAC}
\begin{tabular}{lccc}
\hline
Decay mode & $N_{obs}$ & $\epsilon(\%)$ & Branching fraction  \\
\hline
$\jpsi\rar \ppb\eta,\eta\rar \gamma\gamma$ & $12220\pm 149$
& $28.70$ & $B(\jpsi\rar \ppb\eta)=(1.93\pm 0.02\pm 0.18)\times10^{-3}$ \\
$\jpsi\rar \ppb\eta,\eta\rar \ppp$ & $954\pm 45$
& $4.20$ & $B(\jpsi\rar \ppb\eta)=(1.83\pm 0.09\pm 0.24)\times10^{-3}$ \\
$\jpsi\rar \ppb\etap,\etap\rar \pip\pin\eta$ & $65\pm 12$
& $3.38$ & $B(\jpsi\rar \ppb\etap)=(2.31\pm 0.43\pm 0.34)\times10^{-4}$ \\
$\jpsi\rar \ppb\etap,\etap\rar \gamma\rhoo$ & $200\pm 29$
& $7.48$ & $B(\jpsi\rar \ppb\etap)=(1.85\pm 0.27\pm 0.31)\times10^{-4}$
\\ \hline
\end{tabular}
\end{center}
\end{table*}

\section{\boldmath Systematic errors} \label{SYSE}
In our analysis, the systematic errors on the branching fractions
come from the uncertainties in the MDC tracking, photon
efficiency, particle identification, photon identification, kinematic
fit, background shapes, hadronic interaction model, intermediate decay
branching fraction, the $\pio$ and $\rho$ selection requirements,
intermediate resonance states, and the total number of $\jpsi$ events.
The errors from the different sources are listed in Table~\ref{TABSYS}.

The MDC tracking efficiency has been measured using
$\jpsi\rar\rho\pi$, $\Lambda\bar{\Lambda}$, and
$\psi(2S)\rar\pi^{+}\pi^{-}J/\psi$, $J/\psi$ to $\mu^{+}\mu^{-}$.  The
MC simulation agrees with data within 1 to 2$\%$ for each charged
track~\cite{NIM}. Thus $4\%$ is regarded as the systematic error for
the two charged-track mode, and $8\%$ for the four charged-track final
states.

The photon detection efficiency has been studied using a sample of
$J/\psi \rar\rho\pi$~\cite{NIM} decays; the difference between data
and MC simulation is about $2\%$ for each photon. In this analysis,
$2\%$ is included in the systematic error for one-photon modes and $4\%$
for two-photon modes.

The charged pion PID efficiency has been studied using
$J/\psi\rar\rho\pi$ decays~\cite{NIM}.  The PID efficiency from data
is in good agreement with that from MC simulation with an average
difference that is less than $1\%$ for each charged pion.  Here $2\%$
is taken as the systematic error for identifying two pions.

The proton PID efficiencies have been studied using $\jpsi \rar p
\bar{p}\pip\pin$ decays. The main difference between data and MC
simulation occurs for tracks with momentum less than 0.35 GeV/$c$. We
determine a weighting factor for identifying a proton or anti-proton
as a function of momentum from studies of the $\jpsi\rar
p\bar{p}\pip\pin$ channel.  After considering the weight of each
particle in an event, the difference between data and MC is determined
to be $\frac{\epsilon_{DT}}{\epsilon_{MC}}=0.9739\pm0.0078$ for
$\eta\rar 2\gamma$, $0.9582\pm0.0199$ for $\eta\rar 3\pi$,
$0.8228\pm0.0211$ for $\etap\rar \pip\pin\eta$, and $0.8522\pm0.0140$
for $\etap\rar \gamma\rho^{0}$. We take $f_{1}=0.9739$,
$f_{2}=0.9582$, $f_{3}=0.8228$, and $f_{4}=0.8522$ as efficiency
correction factors for the corresponding decay channel, and $0.8\%$,
$2.1\%$, $2.6\%$, and $1.6\%$ are taken as the errors associated with
identifying protons and anti-protons, respectively. The PID systematic
errors for the four decay modes are listed in Table ~\ref{TABSYS}.

For the systematic error of photon ID, which arises mainly from the
simulation of fake photons, $\ppb$ and $\jpsi \to \ppb\pip\pin$ data
samples were selected and $10^{5}$ simulated $\ppb$ and $\jpsi \to
\ppb\pip\pin$ events were generated, with real and fake photons.  The
decay mode $\jpsi\to \ppb$ is used for the photon ID systematic error
of $\jpsi\to \ppb\eta$ $(\eta\to2\gamma)$, and the decay mode $\jpsi \to
\ppb\pip\pin$ for $\jpsi\to \ppb\eta$ $(\eta\to 3\pi)$ and $\jpsi\to
\ppb\etap$.  From the decay mode $\jpsi\to \ppb$, the fake photon
differences between data and MC is about $2.0\%$, while for the decay
mode $\jpsi \to\ppb\pip\pin$, the difference is $1.6\%$.  Here
$2.0\%$ is taken as the systematic error associated with photon ID for
the decay mode determined to be $\jpsi\to\ppb\eta$
$(\eta\to\gamma\gamma)$, and $1.6\%$ for the decay modes
$\jpsi\to\ppb\eta$ $(\eta\to 3\pi)$ and $\jpsi\to\ppb\etap$.

In Ref.~\cite{PRD7}, the uncertainty of the 4C kinematic fit is $4\%$,
which we include here in the systematic error. The uncertainty of the
5C kinematic fit is $4.1\%$ in Ref.~\cite{RHOP}. Here we
conservatively take $5\%$ as the systematic error from the 5C
kinematic fit for the decay mode $\etap\to \pip\pin\eta$.

The systematic errors of the background uncertainty is obtained by
changing the range of the fit and varying the order of the polynomial
background.  The errors range from 0.8$\%$ to 7.3$\%$ in all decay
modes (see Table~\ref{TABSYS} for detail).

There are two models, FLUKA and GCALOR, used for simulating hadronic
interactions; the different models lead to different detection
efficiencies.  The difference between them is regarded as a systematic
error.  For the decay $\jpsi \to \ppb\eta$ $(\eta \rar 2\gamma)$, the
difference is very small and negligible.  For the other decay modes,
it is about 1.4$\%$ for $\jpsi\rar\ppb\eta$ $(\eta\to\ppp)$,
$\jpsi\rar\ppb\etap$ $(\etap\to\pip\pin\eta)$, and 5.2$\%$ for
$\jpsi\rar\ppb\etap$ $(\etap\to\gamma\rhoo)$.

The branching fractions for the decays $\pio \rar 2\gamma$, $\eta \rar
2\gamma$, $\eta \rar \ppp$, $\etap\rar\pip\pin\eta$, and
$\etap\rar\gamma\rho$ are taken from the PDG~\cite{PDG}. The errors on
these branching fractions are systematic errors in our measurements.

For the $\eta\to 3\pi$ mode, the $\pio$ mass is required to satisfy
$|M_{\gamma\gamma}-M_{\pio}|<0.04$ GeV/$c^{2}$.  To study the
systematic error associated with this requirement, $\pio$ samples are
selected and simulated using $\jpsi \to \rho\pi$, and the data and MC
efficiencies in the 3$\sigma$ signal region are compared with using
the requirement or not, the difference is about $1\%$.  Here it is taken as
the systematic error caused by the $\pio$ requirement.

For the $\etap\to \gamma\rho$ mode, we require that
$|M_{\pip\pin}-M_{\rho}|<0.20$ GeV/$c^{2}$.  According to
Ref.~\cite{FANGP}, the uncertainty associated with this requirement
is $5.9\%$. Here we take this as the systematic error for the $\rho$
mass requirement.

In the signal MC simulation, we assume the presence of N(1440),
N(1535), N(1650), and N(1800) in the $\ppb\eta$ channel.  If some of
these resonances are not included, the efficiency of this channel
changes.  These differences are taken as the systematic error associated
with possible intermediate states.  The total systematic error
associated with this is taken as the sum added in quadrature.  For the
decay modes with an $\etap$, we take the difference in efficiency
determined assuming the decay proceeds via an intermediate $N(2090)$
resonance compared with phase space generation as the systematic error
(see Table~\ref{TABSYS} for detail).

The uncertainty of the total number of $\jpsi$ events is $4.7\%$
\cite{FSS}.  Combining all errors in quadrature gives total systematic
errors of $9.3\%$ for $\eta\rar\gamma\gamma$, $12.9\%$ for
$\eta\rar\ppp$, $14.8\%$ for $\etap\rar\pip\pin\eta$, and $16.6\%$ for
$\etap\rar\gamma\rho$.

\begin{table*}
\begin{center}
\caption{ Summary of systematic errors; ``-'' means no contribution.}
\vspace{0.5cm}
\label{TABSYS}
\begin{tabular}{lcccc}
\hline
Sources
& & Relative error ($\%$) & & \\
\hline
Decay modes & $\eta\rar 2\gamma$ & $\eta\rar \ppp$ & $\etap\rar \pip\pin\eta$
& $\etap\rar \gamma\rhoo$ \\
\hline
MDC tracking & 4 & 8 & 8 & 8 \\
Photon detection efficiency & 4 & 4 & 4 & 2 \\
Particle ID & $\sim$1 & 4.1 & 4.6 & 3.6 \\
Photon ID & 2.0 & 1.6 & 1.6 & 1.6 \\
Kinematic fit &4.0 & 4.0 & 5.0 & 4.0 \\
Background uncertainty & $\sim$1 & 3.1 & 7.3 & 5.8 \\
Hadronic Interaction Model & $\sim$0 & 1.4 & 1.4 & 5.2 \\
Intermediate decay Br. Fr. & $\sim$ 1 & 1.2 & 3.1 & 3.1 \\
$\pio$ selection & - &$\sim$1 & -& \\
$\rho$ selection & - & - & - & 5.9 \\
Intermediate resonances & 3.0 & 4.0 & 2.0 & 7.1 \\
Number of $\jpsi$ events & 4.7 & 4.7 & 4.7 & 4.7 \\
Total systematic error & 9.3 & 12.9 & 14.8 & 16.6 \\
\hline
\end{tabular}
\end{center}
\end{table*}

\section{\boldmath Results}
Table~\ref{TABFAC} shows the branching fractions of the two channels
into their different decay modes; the first error is statistical
and the second is systematic. The results for the different decay
modes in the same channel are consistent
within errors and 
are combined after taking out the common systematic errors (8.37 \%
for the $\eta$ mode and 10.8\% for the $\eta^{'}$ mode):
\begin{eqnarray*}
Br(\jpsi\rar\ppb\eta)=(1.91\pm0.17)\times 10^{-3},\\
Br(\jpsi\rar\ppb\etap)=(2.00\pm0.36)\times 10^{-4}.
\end{eqnarray*}
In comparison with previous measurements of $\jpsi\to\ppb\eta$ and
$\jpsi\to\ppb\etap$, the present results are of higher precision.

 Using the result of $\jpsi\to\ppb\eta$ from this analysis and
that of $\jpsi\to\ppb$ in Ref.~\cite{LXLM},
we determine:
\begin{eqnarray*}
\frac{\Gamma(\jpsi\rar\ppb\eta)}{\Gamma(\jpsi\rar\ppb)}=0.85\pm0.08.
\end{eqnarray*}
This is consistent with the calculation based on the soft-pion
theorem, and indicates that the contribution of $N^{*}$- pole diagram
is dominant for the $\jpsi\to\ppb\eta$ mode. 

\section{\boldmath Acknowledgments}

The BES collaboration thanks the staff of BEPC and computing
center for their hard efforts. This work is supported in part by
the National Natural Science Foundation of China under contracts
Nos. 10491300, 10225524, 10225525, 10425523, 10625524, 10521003, 10821063,
10825524, the Chinese Academy of Sciences under contract No. KJ 95T-03, the
100 Talents Program of CAS under Contract Nos. U-11, U-24, U-25,
and the Knowledge Innovation Project of CAS under Contract Nos.
U-602, U-34 (IHEP), the National Natural Science Foundation of
China under Contract No. 10225522 (Tsinghua University), and the
Department of Energy under Contract No. DE-FG02-04ER41291 (U.
Hawaii).

\end{document}